\documentstyle[pre,aps,epsf]{revtex}

\draft
\begin{document}

\author{D.L.Zhou, P.Zhang and C. P. Sun$^{a,b}$}
\address{Institute of Theoretical Physics, Academia Sinica, \\
P.O.Box 2735, Beijing 100080, China}
\title{High order quantum decoherence via multi-particle amplitude for boson system}
\maketitle

\begin{abstract}
In this paper we depict the high order quantum coherence of a
boson system by using the multi-particle wave amplitude, whose
norm square is just the high order correlation function. This
multi-time amplitude can be shown to be a superposition of several
``multi-particle paths". When the environment or a apparatus
entangles with them to form a generalized ``which-way" measurement
for many particle system, the quantum decoherence happens in the
high order case dynamically. An explicit illustration is also
given with an intracavity system of two modes interacting with a
moving mirror.
\end{abstract}

\pagenumbering{roma}

\thispagestyle{empty} \vspace{15mm}
\widetext \vspace{4mm}

\pagenumbering{arabic} 

\section{Introduction}

A most profound concept in quantum mechanics is coherent
superposition of quantum states. This is obviously reflected by
the special interference of two or more ``paths'' in terms of
single particle wave function. However, this coherence phenomenon
does not sound very marvelous for the same circumstances can also
occur in classical case,such as an optical interference. To
manifest the intrinsically quantum features of coherence beyond
the classical analogue, Glauber's higher order quantum correlation
function (HOQCF) was introduced [1] to accounts for such higher
order quantum coherence effect in the Hanbury-Brown-Twiss
experiment [2].

The quantum coherence both in the first order version and its
higher order one shows the embodying of the wave nature in the
world of the microscopic particles, but it is not robust due to
the wave function collapse or quantum decoherence caused by a
quantum measurement or by the coupling environment. Roughly, in
the single particle picture, this decoherence phenomenon losing
coherence can be understood according to the quantum entanglement
of the considered system with the environment or the measuring
apparatus. Obviously, this entanglement implies a ``which-path''
detection[3]. Precisely speaking, in a initial coherent
superposition $|\psi _s\rangle =\sum c_n|n\rangle $ , each system
state $|n\rangle $ corresponding a ``path'' is correlated with an
environment state $|e_n\rangle $ to form an
entangling state $|\psi _T\rangle =\sum c_n|n\rangle \bigotimes |e_n\rangle $%
. The existence of the different states $|e_n\rangle $ distinguish among the
``paths''of different $|n\rangle $. The interferences terms in $%
I(x)=Tr(\langle x|\psi _T\rangle \langle \psi _T|x\rangle )$ can disappear
when the environment state $|e_n\rangle $ are very distinct, i.e., each path
is explicitly labelled $\langle e_m|e_n\rangle =\delta _{mn}$.

For the first order decoherence, the above well-known explanation
in terms of ``wich-path'' detection mechanism is more simple, but
very profound. However, it is not obvious yet whether this
mechanism can account for decoherence in the higher effects: This
is because we do not exactly know what is the ``paths'' and the
corresponding ``which-path'' detection in the high order version.
More recently, we have touched the quantum decoherence problem in
higher order case [4]. The concrete calculation motivated us to
further consider the ``which-path'' picture for the higher order
quantum decoherence. In this paper, for a boson system we
introduce the concept of the multi-particle wave amplitude, whose
norm square is just the high order correlation function. As an
effective wave function, this multi-time amplitude can be shown to
be a supposition of several components. When the environment or a
apparatus entangles with them, the quantum decoherence occurs in
the high order case dynamically. This decoherence process losing
the higher order coherence can be explained as a generalized
``which-path'' measurement for the defined multi-particle paths.

\section{Multi-Particle Paths for Probability Amplitudes}

Firstly, we start with considering the meaning of the ``path'' in
the high order quantum correlation. The typical example of the
higher order quantum coherence [5] is that the single-component
state $|1_k,1_{k^{\prime }}\rangle $ of the two independent
photons with momenta $k$ and $k^{\prime }$ shows
its quantum coherence in its second order quantum correlation function $%
G^{(2)}\equiv G^{(2)}(r_1,r_2,t_1,t_2)$. It can just be written as
of form of the norm square $G^{(2)}=|\psi |^2$ of the equivalent
``two-time wave function''
\begin{equation}
\psi \equiv \psi (r_1,r_2,t_1,t_2)=\langle
00|E^{+}(r_2,t_2)E^{+}(r_1,t_1)|1_k,1_{k^{\prime }}\rangle
\end{equation}
It was also called the biphoton wave packet for the photon field $%
E^{+}(r,t)$ [6]. Especially, we remark that $\psi $ is a coherent
superposition of several ``probability amplitudes''. This result
can be promoted to the universal case with quantum systems of
identical particles.

To see the main physical ideas implied by the above well-known
result, without loss of the generality, we define a ``measuring''
operator of two modes $V$ and $H$ [4]
\begin{equation}
\hat{\phi}=c_V\hat{b}_Ve^{-i\omega _Vt}+c_H\hat{b}_He^{-i\omega _Ht}\equiv
c_V(t)\hat{b}_V+c_H(t)\hat{b}_H
\end{equation}
where $b_H$ and $b_V$ are the annihilation operators of the boson
system. The generalized second order correlation function[5]
\begin{eqnarray}
\hat{G}^{(2)} &=&\langle 1_V1_H|\hat{\phi}^{\dagger }(t_1)\hat{\phi}%
^{\dagger }(t_2)\hat{\phi}(t_2)\hat{\phi}(t_1)|1_V,1_H\rangle  \nonumber \\
\mbox{} &=&|\langle 0,0|\hat{\phi}(t_2)\hat{\phi}(t_1)|1_V,1_H\rangle
|^2\equiv |\Psi (t_1,t_2)|^2
\end{eqnarray}
The two time wave function $\Psi (t_1,t_2)$ can be understood in
terms of the two ``paths'' picture from the initial state
$|1_V,1_H\rangle $ to the finial one $|0,0\rangle $: \vskip 5mm

\begin{center}
\begin{tabular}{|c|}
\hline
$
\begin{array}{ccccc}
|1_V,1_H\rangle & \stackrel{c_H(t_1)}{\longrightarrow } &
|1_V,0_H\rangle &
\stackrel{c_{_V}(t_2)}{\longrightarrow } & |0,0\rangle \\
& \searrow \stackrel{c_V(t_1)}{}&  & \stackrel{c_H(t_2)}{}\nearrow &  \\
&  & |0_V,1_H\rangle&  &
\end{array}
$
\\
\hline
\end{tabular}
\end{center}

They are just associated with the two amplitudes forming a coherent
superposition

\begin{equation}
\Psi (t_1,t_2)=c_Vc_He^{-i\omega _Vt_2-i\omega _Ht_1}+c_Hc_Ve^{-i\omega
_{_H}t_2-i\omega _{_V}t_1}
\end{equation}
Correspondingly, the second order correlation function
\begin{equation}
G^{(2)}=2|c_Vc_H|^2[1+\cos ([\omega _V-\omega _H][t_2-t_1])]
\end{equation}

The above observation for the second order quantum coherence can also be
discovered in the higher order case. Our arguments in this paper will be
based on two novel observations: $a.$ The equivalent field operator $\hat{%
\Phi}=\sum c_n\hat{b}_n$ is specified for a quantum measurement to
a superposition state $|\phi \rangle =\sum c_n|n\rangle $. $b.$
For a certain initial single component state $|s_0\rangle $ of $N$
particles system, the $n $'th order quantum correlation function
\begin{equation}
G^{(n)}(r_1,r_2,\cdots ,r_n,t_1,t_2,\cdots ,t_n)=|\psi ^{(n)}|^2
\end{equation}
can be written as the norm square of an effective wave function $\psi ^{(n)}$%
, which is just a superposition of many amplitudes.

For the seek of simplicity, we only consider the third order situation when
the initial state is in the state $|1_H,2_V\rangle $. Indeed, the
generalized third order correlation function
\begin{eqnarray}
\hat{G}^{[3]} &=&\langle 2_V1_H|\hat{\phi}^{\dagger }(t_1)\hat{\phi}%
^{\dagger }(t_2)\hat{\psi}^{\dagger }(t_3)\hat{\psi}(t_3)\hat{\phi}(t_2)\hat{%
\phi}(t_1)|2_V,1_H\rangle   \nonumber \\
\mbox{} &=&|\langle 0,0|\hat{\psi}(t_3)\hat{\phi}(t_2)\hat{\phi}%
(t_1)|2_V,1_H\rangle |^2\equiv |\Psi (t_1,t_2,t_3)|^2
\end{eqnarray}
is indeed a norm square of the two time wave function $:$

\begin{eqnarray}
\Psi (t_1,t_2,t_3) &=&\sqrt{2}{c_V}^2c_He^{-i\omega _V(t_3+t_2)-i\omega
_Ht_1}+  \nonumber \\
&&\sqrt{2}c_H{c_V}^2e^{-i\omega _Ht_2-i\omega _V(t_3+t_1)}+\sqrt{2}c_H{c_V}%
^2e^{-i\omega _Ht_3-i\omega _V(t_2+t_1)}
\end{eqnarray}
Each term in the above equivalent wavefunction is contributed by
the corresponding one of  the four
``paths'' from $|2_V,1_H\rangle $ into $|0,0\rangle $:\vspace{%
5mm}

\begin{center}
\begin{tabular}{|l|}
\hline
\begin{tabular}{ccccccc}
&  & $|2_V,0_H\rangle $ & $\stackrel{c_{_V}(t_2)}{\longrightarrow }$ & $%
|1_V,0_H\rangle $ &  &  \\
& $\nearrow \stackrel{c_{_H}(t_1)}{}$ &  &  &  & $\stackrel{c_V(t_3)}{}%
\searrow $ &  \\
$|2_V,1_H\rangle $ & $\stackrel{c_{_V}(t_1)}{\longrightarrow }$ & $%
|1_V,1_H\rangle $ & $\stackrel{c_H(t_2)}{\longrightarrow }$ & $%
|1_V,0_H\rangle $ & $\stackrel{c_V(t_3)}{\longrightarrow }$ & $|0,0\rangle $
\\
&  &  & $\searrow \stackrel{c_V(t_2)}{}$ &  & $\stackrel{c_{_H}(t_3)}{}%
\nearrow $ &  \\
&  &  &  & $|0_V,1_H\rangle $ &  &
\end{tabular}
\\ \hline
\end{tabular}
\end{center}

From the above equivalent $3$-time-wave function, the third order
correlation function is explicitly written down
\begin{eqnarray}
G^{(3)} &=&4|{c_V}^2c_H|^2(\frac 32+\cos [(\omega _V-\omega _H)(t_2-t_1)]+ \\
&&\cos [(\omega _V-\omega _H)(t_3-t_1)]+\cos [(\omega _V-\omega
_H)(t_2-t_3)])  \nonumber
\end{eqnarray}
and shows the quantum interference in the time-domain.

From the above calculations for the second and third order quantum
decoherence, the observation can be made that, for a
specially-given initial states, a high order correlation function
may be explicitly written down as the norm square of the
equivalent multi-time-wave function, which is a coherent
superposition of several complex components in associated with the
generalized many-particle paths. It is pointed out that this kind
of many-particle path is not a simple-product of single-particle
paths and it can be determined by the concrete measurement.

\section{Intracavity Model}

With the above introduced concept of``many-particle path", we can
discuss the higher order decoherence problem by considering the
many-particle ``which -path'' measurement. Let us use the
following model to sketch this central idea. Our model is
consisted of non-dissipative bosons in two modes. The problem is
studied via calculating the second order quantum correlation
functions in the Heisenberg picture. We take $\hbar =1$ in this
paper.

The model Hamiltonian $\hat{H}=\hat{H}_0+\hat{V}$ is defined by
\begin{eqnarray}
\hat{H}_0 &=&\omega _V\hat{b}_V^{\dagger }\hat{b}_V,  \nonumber \\
\hat{V} &=&\sum_j\omega _j\hat{a}_j^{\dagger }\hat{a}_j+\sum_j[d_V(\omega _j)%
\hat{b}_V^{\dagger }\hat{b}_V+d_H(\omega _j)\hat{b}_H^{\dagger }\hat{b}_H](%
\hat{a}_j^{\dagger }+\hat{a}_j),
\end{eqnarray}
where $\hat{H}_0$ is the free Hamiltonian of the system, $\hat{V}$ the free
Hamiltonian of the reservoir (or a detector) plus the interaction between
the system and the reservoir; and $\hat{b}_V^{\dagger }(\hat{b}_V),\hat{b}%
_H^{\dagger }(\hat{b}_H)$ the creation (annihilation) operators for two
modes with frequencies $\omega _V$ and $\omega _H=0$ . The operators $\hat{a%
}_j^{\dagger }(\hat{a}_j)$ are creation (annihilation) operators
of the reservoir modes of frequencies $\omega _j$ . The
frequency-dependent constant $d_H(\omega _j)$ ($d_V(\omega _j)$)
measures the coupling constant between $H$($V$ )mode and the $j$
mode of the reservoir. The most important feature of the model is
that $[H_0,V]=0$, i.e. the system does not dissipate energy to the
reservoir, but it can leave imprinting on the reservoir since, for
different number states $|n_V,n_H\rangle $, there are different
interactions $\sum_j[n_Vd_V(\omega _j)+n_Hd_H(\omega
_j)](\hat{a}_j^{\dagger }+\hat{a}_j)$ acting on the oscillator
reservoir with different driving forces $\sim n_Vd_V(\omega
_j)+n_Hd_H(\omega _j)$.

When there is only one mode in external system we can physically
realized this model as an intracavity model[4,8]: Two mode cavity
field interact with a moving wall of the cavity, which is attached
to a spring and can be regarded as a harmonic oscillator with a
small mass. The coupling of fields to the cavity wall (a moving
mirror) is just by the radiation pressure forces
proportioned to the photon numbers by $\hat{b}_H^{\dagger }\hat{b}_H$ and $\hat{%
b}_V^{\dagger }\hat{b}_V$.

The second order coherence is directly determined by the second order
correlation.
\begin{equation}
G[t,t^{\prime },\hat{\rho}(0)]=Tr(\hat{\rho}(0)\hat{B}^{\dagger }(t)\hat{B}%
^{\dagger }(t^{\prime })\hat{B}(t^{\prime })\hat{B}(t)),
\end{equation}
which is defined as a functional of the density operator $\hat{\rho}_S(0)$
of the whole system at a given time $0$. Here, the bosonic field operator
\begin{eqnarray}
\hat{B}(t) &=&\exp (i\hat{H}t)[c_1\hat{b}_H+c_2\hat{b}_V]\exp (-i\hat{H}t)=
\\
&&\exp (i\hat{V}t)[c_1\hat{b}_H+c_2\hat{b}_V\exp (-i\omega _Vt)]\exp (-i\hat{%
V}t)
\end{eqnarray}
describes a specific quantum measurement[4] with respect the
polarized photon
states $|+\rangle =$ $c_1|H\rangle $ $+c_2|V\rangle $ and $|-\rangle =$ $%
c_2|H\rangle $ $-c_1|V\rangle $ where $c_1$ and $c_2$ satisfy the
normalization relation $|c_1|^2+|c_2|^2=1$. Without loss of the generality,
we take $c_1=c_2=1/\sqrt{2}$ standing for a given measurement as follows.

To examine whether the macroscopic feature of the reservoir causes the
second order decoherence or not , we consider the whole system in an initial
state
\begin{equation}
|\psi (0)\rangle =|1_H,1_V\rangle \otimes |\{0_j\}\rangle ,
\end{equation}
where $|\{0_j\}\rangle $ is the vacuum state of the reservoir.
Here, we have denoted the general Fock states of the many mode
field by $|\{n_j\}\rangle \equiv |n_1,n_2,...\rangle $.

For the total system, in stead of defining the equivalent
``two-time wave function'' in the above section , we define an
effective two-time state vector
\begin{equation}
|\psi _B(t,t^{\prime })\rangle =\hat{B}(t^{\prime })\hat{B}(t)|\psi
(0)\rangle .
\end{equation}
to re-write the second order correlation function as
\begin{equation}
G[t,t^{\prime },\hat{\rho}(0)]=\langle \psi _B(t,t^{\prime })|\psi
_B(t,t^{\prime })\rangle
\end{equation}
It is interested that the effective state vector can be evaluated as the
superposition
\begin{eqnarray}
|\psi _B(t,t^{\prime })\rangle  &=&\frac 12e^{i\hat{V}(0,0)t^{\prime }}[\exp
(-i\omega _Vt^{\prime })e^{-i\hat{V}(1,0)t^{\prime }}e^{i\hat{V}(1,0)t}+ \\
&&\exp (-i\omega _Vt)e^{i\hat{V}(0,0)t^{\prime }}e^{-i\hat{V}(0,1)t^{\prime
}}e^{i\hat{V}(0,1)t}]e^{-i\hat{V}(1,1)t}|\{0_j\}\rangle \otimes
|0_H,0_V\rangle
\end{eqnarray}
of two components for the two paths from the initial two particle state $%
|1_H,1_V\rangle $ to the two particle vacuum $|0_H,0_V\rangle $ . It should
be noticed that the effective actions of the reservoir
\begin{equation}
\hat{V}(m,n)\equiv \sum_j\hat{V}_j(m,n)=\sum_j\omega _j\hat{a}_j^{\dagger }%
\hat{a}_j+\sum_j(d_V(\omega _j)m+d_H(\omega _j)n)(\hat{a}_j^{\dagger }+\hat{a%
}_j)
\end{equation}
can label the different paths and thus lead to the higher order
quantum decoherence. The above result clearly demonstrates that,
in presence of the reservoir, the different probability amplitudes
($\sim $ $\exp (-i\omega
_Vt^{\prime })$ and $\exp (-i\omega _Vt)$) from $|1_H,1_V\rangle $ to $%
|0_H,0_V\rangle $ entangle with the different states ($\frac 12e^{i\hat{V}%
(0,0)t^{\prime }}e^{-i\hat{V}(1,0)t^{\prime }}e^{i\hat{V}(1,0)t}e^{-i\hat{V}%
(1,1)t}|\{0_j\}\rangle $ and $\frac 12e^{i\hat{V}(0,0)t^{\prime }}e^{-i\hat{V%
}(0,1)t^{\prime
}}e^{i\hat{V}(0,1)t}e^{-i\hat{V}(1,1)t}|\{0_j\}\rangle $ ) of the
reservoir. This is just physical source of the higher order
quantum decoherence. In the following section an explicit
calculation of the second order correlation function will be given
to show this crucial observation.

\section{Dynamic Decoherence in Higher Order Case}

In our calculation, the second order correlation function
\begin{equation}
G[t,t^{\prime },\hat{\rho}(0)]=\frac 12+\frac 14e^{i\omega _V(t-t^{\prime
})}\prod_jF_j+\frac 14e^{-i\omega _V(t-t^{\prime })}\prod_jF_j^{*},
\end{equation}
is firstly expressed as a factorization form[9] where each factor
\begin{equation}
F_j=\langle 0_j|\hat{u}_j^6(t_6)|0_j\rangle \equiv \langle 0_j|e^{i\hat{V}%
_j(1,1)t}e^{-i\hat{V}_j(0,1)t}e^{i\hat{V}_j(0,1)t^{\prime }}e^{-i\hat{V}%
_j(1,0)t^{\prime }}e^{i\hat{V}_j(1,0)t}e^{-i\hat{V}_j(1,1)t}|0_j\rangle
\end{equation}
is a two-time transition amplitude of the $j^{\prime }th$ mode of
the reservoir. Obviously, the term $\prod_jF_j$ determines the
extent of coherence and decoherence in the second order case,
which is called ``the decoherence factor'' and plays the role just
as the same as that in the first order decoherence[9].

In the following, to given the factor $F_j$ explicitly , we adopt
the Wei-Norman method [10,11] to calculate the equivalent time
evolution defined by $\hat{u}_j^6(t_6)$. It can be image as an
evolution governed by a discrete
time-dependent Hamiltonian $H(t)$ dominated by $\hat{V}_j(1,1),-\hat{V}_j(1,0),\hat{V}%
_j(1,0),-\hat{V}_j(0,1),\hat{V}_j(0,1)$ and $-\hat{V}_j(1,1)$in six time-intervals $%
[t_0=0,t_1=t],[t_1,t_2=2t],[t_2,t_3=2t+t^{\prime
}],[t_3,t_4=2t+2t^{\prime }],[t_4,t_5=3t+2t^{\prime }],$
$[t_5,t_6=4t+2t^{\prime }]$ respectively. In the $k$-th step of
calculation, we take the final state of ($k-1)$-th step as its
initial state. Therefore,we obtain $\hat{u}_j^6(t_6)$ as the sixth
step evolution

\begin{equation}
\hat{u}_j^6(t_6)=e^{{g_{1j}^6(t_6)\hat{a}_j}^{\dagger }}e^{{g_{2j}^6(t_6)\hat{a}_j}%
^{\dagger
}\hat{a}_j}e^{{g_{3j}^6(t_6)}\hat{a}_j}e^{{g_{4j}^6(t_6)}}
\end{equation}
Here, ${g_{kj}^6(t_6)(k=1,2,3,4)}$ are the coefficients that can
be explicitly
obtained, but for the calculation of the $j$-th component $F_j=\exp [{%
g_{4j}^6}(t_6)]$ of the decoherence factor$,$ we only need to know
${g_{4j}^6.}$ The detailed discussion in the appendix gives

\begin{eqnarray}
{g_{4j}^6}(t_6) &=&-2\frac{[d_H(j)-d_V(j)]^2}{\omega _j^2}\sin
^2[\frac{\omega
_j(t^{\prime }-t)}2]+i\frac{d_V^2(j)-d_H^2(j)}{\omega _j^2} \\
&\mbox{}&[\omega _j(t^{\prime }-t)+2(1-\cos (\omega _j[t^{\prime }-t])\sin
\omega _jt]+(1-2\cos \omega _jt)\sin (\omega _j[t^{\prime }-t])].  \nonumber
\end{eqnarray}
Since the real part of $g_{4j}^6$ is no more than zero, the norm
of $F_j$ is no more than one. From the same argument as in the
first order decoherence [9], the universal factorization structure
of the decoherence factor implies the second order decoherence in
the macroscopic limit.

In order to demonstrate the above arguments quantitatively, we
give the numerical results for the second order decoherence for
different numbers N of the quantum oscillators. As N increase,
these results are illustrated in FIG.1. In the numerical
calculation, the coupling constants $\{d_V(j)\}$
take random values in the domain $[0.8,1.0]$, the coupling constants $%
\{d_H(j)\}$ in $[0.2,0.4]$, and the frequencies $\{\omega _j\}$ in $[0.5,1.5]
$. The other parameters is given in the caption of the figure. \vskip 0.2cm
\begin{figure}[tbp]
\begin{tabbing}
\= \epsfxsize=8.3cm\epsffile{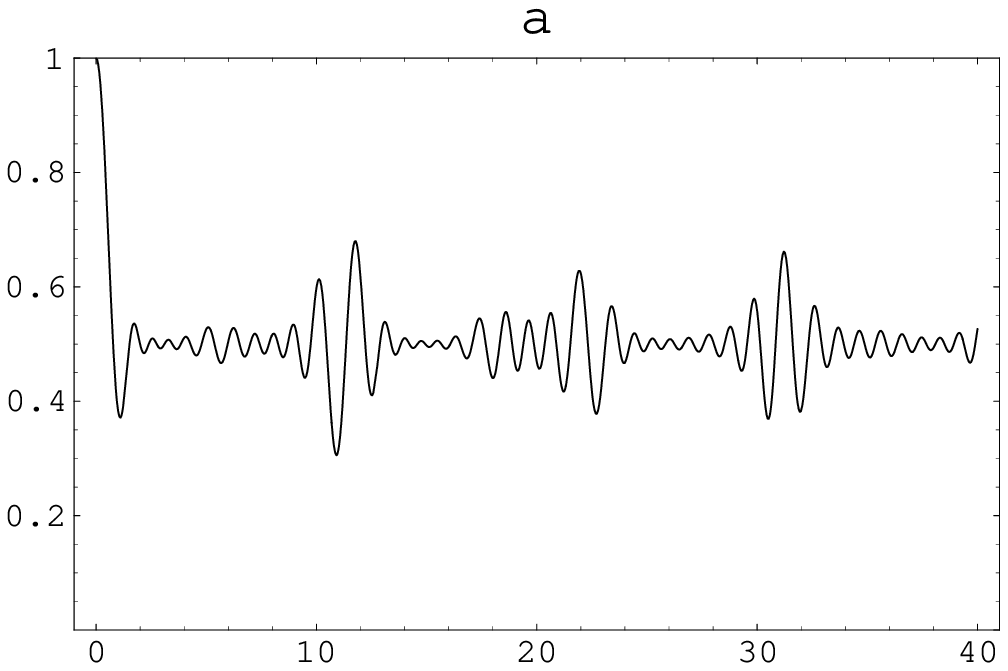}
\= \epsfxsize=8.3cm\epsffile{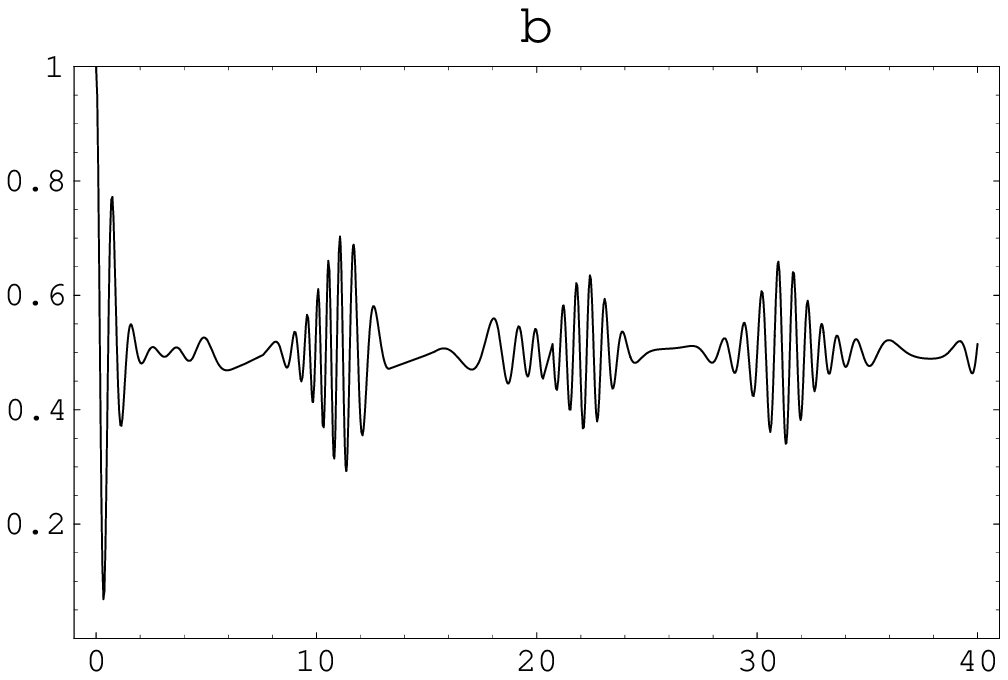}\\
\> \epsfxsize=8.3cm\epsffile{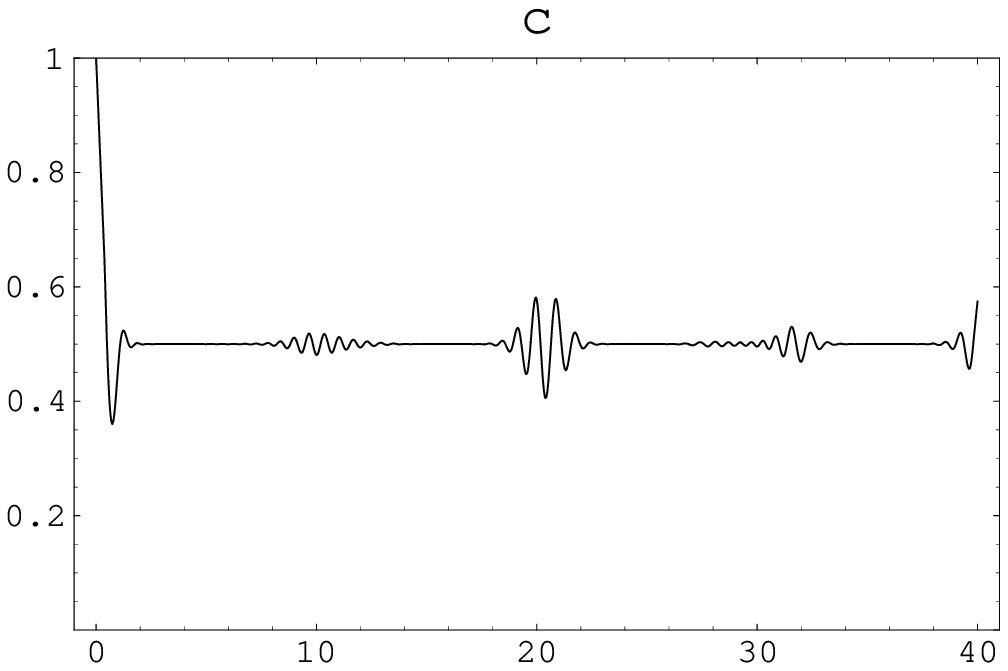}
\> \epsfxsize=8.3cm\epsffile{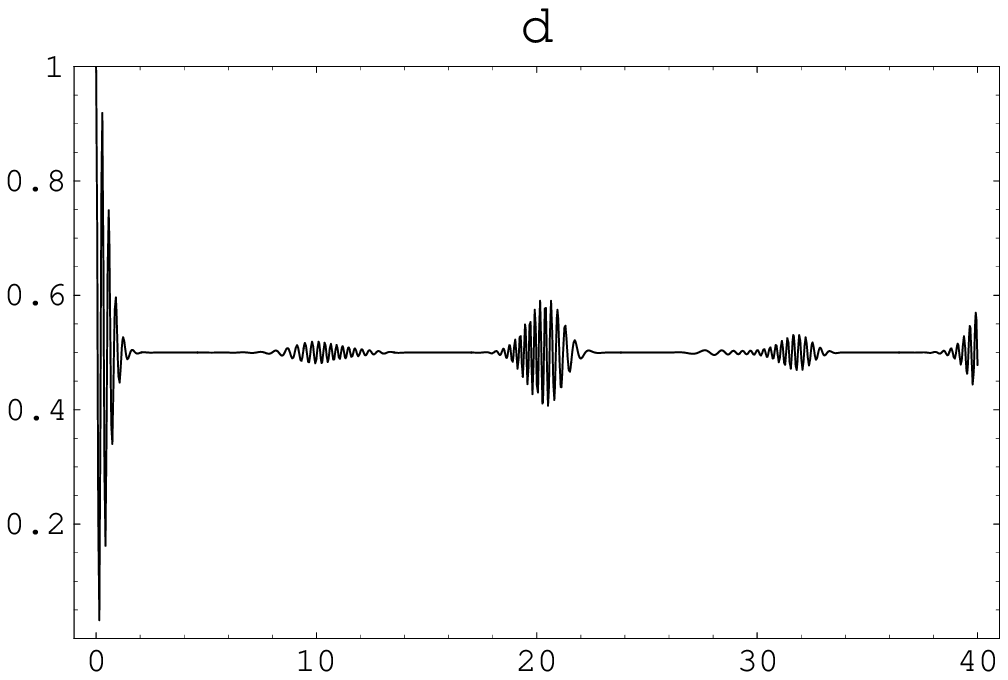}
\end{tabbing}
\vskip 0.2cm \caption{The horizontal axe denotes time period
$t^{\prime }-t$, the vertical axe denotes the second order
correlation function $G[t,t^{\prime },\hat{\rho}(0)]$,
parameters ${\omega _V=1.0}$, (a)${N=5,t=0}$, (b)${N=5,t=5}$, (c)${N=10,t=0}$%
, (d)${N=10,t=5}$.}
\end{figure}

From the above illustration, we see clearly that the second order
correlation is not only a function of the time period $t^{\prime
}-t$, but also a function of time $t$. The amplitude of the second
order correlation function is mainly determined by the time
period, but the phase is determined by the two parameters in the
same time. More important, with the increasing of the number of
the quantum oscillators, the second order coherence vanishes
faster and faster, and the quantum revival amplitude becomes
smaller. With extrapolation, it can be expected that, when the
number of the quantum oscillators limit to infinity, i.e., in the
macroscopic limit of the reservoir, the second order coherence
will decoherence in a short time and no quantum revival can be
observed.

\appendix
\section*{\mbox{}}

In this appendix, the Wei-Norman method [10,11] is adopted to
calculate the second order decoherence factor $F_j$. The
calculation is completed in six steps. During the time period
$[t_{k-1},t_k]$, the evolution for $W_j^k(t,t^{\prime })$ is
dominated by the Hamiltonian
\begin{equation}
\hat{h}_j^k={\alpha _j^k}\hat{a}_j^{\dagger }\hat{a}_j+{\beta _j^k}\hat{a}%
_j^{\dagger }+{\gamma _j^k}\hat{a}_j,\hspace{0.5cm}\{k=1,2,\cdots
,6\}.
\end{equation}
The coefficients $\alpha _j^k,\beta _j^k,\gamma _j^k$ and the
times intervals $T_k=t_k-t_{k-1}$ take different values in
different steps:
\begin{eqnarray}
&&\alpha _j^1=\omega _j,\beta _j^1=\gamma _j^1=d_V(\omega _j)+d_H(\omega
_j),T_1=t,  \nonumber \\
&&\alpha _j^2=-\omega _j,\beta _j^2=\gamma _j^2=d_V(\omega _j),T_2=t,
\nonumber \\
&&\alpha _j^3=\omega _j,\beta _j^3=\gamma _j^3=d_V(\omega _j),T_3=t^{\prime
},  \nonumber \\
&&\alpha _j^4=-\omega _j,\beta _j^4=\gamma _j^4=-d_H(\omega
_j),T_4=t^{\prime },  \nonumber \\
&&\alpha _j^5=\omega _j,\beta _j^5=\gamma _j^5=d_H(\omega _j),T_5=t,
\nonumber \\
&&\alpha _j^6=\omega _j,\beta _j^6=\gamma _j^6=-d_V(\omega _j)-d_H(\omega
_j),T_6=t.
\end{eqnarray}
Due to the fact that ${\hat{a}_j}^{\dagger }\hat{a}_j$, ${\hat{a}_j}%
^{\dagger }$, $\hat{a}_j$, $1$ form a closed algebra - the Heisenberg-Wely
algebra, the unitary time evolution operator at every step takes the
following form
\begin{equation}
\hat{u}_j^k(T)=e^{{g_{1j}^k}(T){\hat{a}_j}^{\dagger }}e^{{g_{2j}^k}(T){\hat{a%
}_j}^{\dagger }\hat{a}_j}e^{{g_{3j}^k}(T)\hat{a}_j}e^{{g_{4j}^k}(T)}.
\end{equation}
for $T\in[t_{k-1},t_k]$ in a special sequence. The benefit of the
above form is that only the coefficient ${g_{4j}^k}(T)$ is needed
to be known in the calculation of the average value of the vacuum
state.

According to the Schr$\ddot{o}$dinger equation $i\frac d{dT}\hat{u}_j^k=\hat{h}%
_j^k\hat{u}_j^k$, the coefficients satisfy the equations:
\begin{eqnarray}
\frac d{dT}{g_{2j}^k} &=&-i{\alpha _j^k},  \nonumber \\
\frac d{dT}{g_{1j}^k}-{g_{1j}^k}\frac d{dT}{g_{2j}^k} &=&-i{\beta _j^k},
\nonumber \\
e^{-{g_{2j}^k}}\frac d{dT}{g_{3j}^k} &=&-i{\gamma _j^k} \\
\frac d{dT}{g_{4j}^k}-{g_{1j}^k}e^{-{g_{2j}^k}}\frac d{dT}{g_{3j}^k} &=&0
\nonumber
\end{eqnarray}
Using the results
\begin{equation}
\frac d{dT}{g_{1j}^k}=-i{\alpha _j^k}{g_{1j}^k}-i{\beta _j^k},\frac d{dT}{%
g_{4j}^k}=-i{\gamma _j^k}{g_{1j}^k}
\end{equation}
obtained by simplifying the system of equations , we get the
solution
\begin{eqnarray}
{g_{1j}^k}(T) &=&({g_{1j}^k}(t_{k-1})+\frac{\beta _j^k}{\alpha
^k})e^{-i{\alpha
_j^k}(T-t_{k-1}}-\frac{\beta _j^k}{\alpha _j^k},  \nonumber \\
{g_{4j}^k}(T) &=&{g_{4j}^k}(t_{k-1})+\frac{\gamma _j^k}{\alpha _j^k}({g_{1j}^k}(t_{k-1})+%
\frac{\beta _j^k}{\alpha _j^k})(e^{-i{\alpha _j^k}(T-t{k-1})}-1)+i\frac{{\beta _j^k}{%
\gamma _j^k}(T-t_{k-1})}{\alpha _j^k}
\end{eqnarray}
Notice that we have used the step-initial conditions
\begin{eqnarray}
g_{1j}^k(t_{k-1}) &=&g_{1j}^{k-1}(t_{k-1}), \\
g_{4j}^k(t_{k-1}) &=&g_{4j}^{k-1}(t_{k-1})
\end{eqnarray}
and the initial conditions $g_{1j}^0(t_0)=g_{4j}^0(t_0)=0$.  Then
we obtain a set of iteration equations

\begin{eqnarray}
g_{1j}^k(t_k) &=&({g_{1j}^{k-1}}(t_{k-1})+\frac{\beta _j^k}{\alpha _j^k})e^{-i{\alpha ^k}%
T_k}-\frac{\beta _j^k}{\alpha _j^k},  \nonumber \\
g_{4j}^k(t_k) &=&{g_{4j}^{k-1}(t_k-1)}+\frac{\gamma _j^k}{\alpha _j^k}({g_{1j}^{k-1}(t_{k-1}}+\frac{%
\beta _j^k}{\alpha _j^k})(e^{-i{\alpha _j^k}T_k}-1)+i\frac{{\beta
_j^k}{\gamma _j^k}T_k}{\alpha ^k}.
\end{eqnarray}

Iterating six times with different initial conditions and
coefficients, the final result of ${g_4^6(t_6)}$ is obtained as
the equation (23).

\vskip 0.2cm {\bf Acknowledgement} This work is supported by the
NSF of China and the knowledged Innovation Programme(KIP) of the
Chinese Academy of Science.


\end{document}